\PassOptionsToPackage{unicode}{hyperref}
\PassOptionsToPackage{hyphens}{url}
\PassOptionsToPackage{dvipsnames,svgnames,x11names}{xcolor}
\documentclass[
  10pt,
  a4paper,
]{article}
\usepackage{xcolor}
\usepackage[margin=22mm]{geometry}
\usepackage{amsmath,amssymb}
\usepackage{iftex}
\ifPDFTeX
  \usepackage[T1]{fontenc}
  \usepackage[utf8]{inputenc}
  \usepackage{textcomp} % provide euro and other symbols
\else % if luatex or xetex
  \usepackage{unicode-math} % this also loads fontspec
  \defaultfontfeatures{Scale=MatchLowercase}
  \defaultfontfeatures[\rmfamily]{Ligatures=TeX,Scale=1}
\fi
\usepackage{lmodern}
\ifPDFTeX\else
\fi
\IfFileExists{upquote.sty}{\usepackage{upquote}}{}
\IfFileExists{microtype.sty}{% use microtype if available
  \usepackage[]{microtype}
  \UseMicrotypeSet[protrusion]{basicmath} % disable protrusion for tt fonts
}{}
\makeatletter
\@ifundefined{KOMAClassName}{% if non-KOMA class
  \IfFileExists{parskip.sty}{%
    \usepackage{parskip}
  }{% else
    \setlength{\parindent}{0pt}
    \setlength{\parskip}{6pt plus 2pt minus 1pt}}
}{% if KOMA class
  \KOMAoptions{parskip=half}}
\makeatother
\usepackage{longtable,booktabs,array}
\usepackage{caption}
\usepackage{calc} % for calculating minipage widths
\usepackage{etoolbox}
\makeatletter
\patchcmd\longtable{\par}{\if@noskipsec\mbox{}\fi\par}{}{}
\makeatother
\IfFileExists{footnotehyper.sty}{\usepackage{footnotehyper}}{\usepackage{footnote}}
\makesavenoteenv{longtable}
\usepackage{graphicx}
\makeatletter
\newsavebox\pandoc@box
\newcommand*\pandocbounded[1]{% scales image to fit in text height/width
  \sbox\pandoc@box{#1}%
  \Gscale@div\@tempa{\textheight}{\dimexpr\ht\pandoc@box+\dp\pandoc@box\relax}%
  \Gscale@div\@tempb{\linewidth}{\wd\pandoc@box}%
  \ifdim\@tempb\p@<\@tempa\p@\let\@tempa\@tempb\fi% select the smaller of both
  \ifdim\@tempa\p@<\p@\scalebox{\@tempa}{\usebox\pandoc@box}%
  \else\usebox{\pandoc@box}%
  \fi%
}
\def\fps@figure{htbp}
\makeatother
\NewDocumentCommand\citeproctext{}{}

\makeatletter
 \let\@cite@ofmt\@firstofone
 \def\@biblabel#1{}
 \def\@cite#1#2{{#1\if@tempswa , #2\fi}}
\makeatother
\newlength{\cslhangindent}
\newlength{\csllabelwidth}
\newenvironment{CSLReferences}[2] % #1 hanging-indent, #2 entry-spacing
 {\begin{list}{}{%
  \setlength{\itemindent}{0pt}
  \setlength{\leftmargin}{0pt}
  \setlength{\parsep}{0pt}
  \ifodd #1
   \setlength{\leftmargin}{\cslhangindent}
   \setlength{\itemindent}{-1\cslhangindent}
  \fi
  \setlength{\itemsep}{#2\baselineskip}}}
 {\end{list}}
\usepackage{calc}

\newcommand{\CSLLeftMargin}[1]{\parbox[t]{\csllabelwidth}{\strut#1\strut}}
\newcommand{\CSLRightInline}[1]{\parbox[t]{\linewidth - \csllabelwidth}{\strut#1\strut}}

\usepackage{microtype}
\usepackage{xurl}
\usepackage{fancyhdr}
\usepackage{enumitem}
\usepackage{caption}
\setlist{nosep}
\usepackage{titlesec}
\titleformat{\section}{\sffamily\large\bfseries}{}{0pt}{}
\titleformat{\subsection}{\sffamily\normalsize\bfseries}{}{0pt}{}
\titleformat{\subsubsection}{\sffamily\small\bfseries\itshape}{}{0pt}{}
\titlespacing*{\section}{0pt}{1.6ex plus 0.4ex minus 0.2ex}{0.8ex plus 0.2ex}
\usepackage{bookmark}
\IfFileExists{xurl.sty}{\usepackage{xurl}}{} % add URL line breaks if available
\makeatletter
\@ifundefined{xmpquote}{}{}
\makeatother
\hypersetup{
  pdftitle={ProteoEM: probabilistic protein abundance estimation from iterative affinity traces},
  colorlinks=true,
  linkcolor={blue},
  filecolor={Maroon},
  citecolor={Blue},
  urlcolor={blue},
  pdfcreator={LaTeX via pandoc}}

\title{ProteoEM: probabilistic protein abundance estimation from
iterative affinity traces}
\author{}
\date{}

\begin{document}
\maketitle

\begin{center}
{\large Narayanan Raghupathy}\\[3pt]
Tensoromics LLC\\[3pt]
\href{mailto:narayanan.raghupathy@tensoromics.com}{narayanan.raghupathy@tensoromics.com}
\end{center}

\section{Abstract}\label{abstract}

Single-molecule affinity mapping enables measurement of individual
proteins and proteoforms, but imperfect and nonspecific probe binding
means that affinity traces may be compatible with multiple proteoforms.
Accurate abundance estimation therefore requires probabilistically
weighting ambiguous traces rather than assigning each trace to a single
candidate.

We developed ProteoEM, an expectation-maximization framework for
weighted proteoform quantification, inspired by transcript abundance
estimation methods for RNA-seq and released as an open-source Python
package. ProteoEM evaluates each observed affinity trace against all
candidate proteoforms using fixed, pre-calibrated probe-response rates
that are separate from abundance estimation. It estimates the abundance
of each proteoform and reports proteoforms that the probes cannot tell
apart as a single group. Because some proteoforms are observed more
readily than others, ProteoEM also corrects for these observation yields
to estimate the composition of the source sample. We show that grouping
traces that carry the same evidence into equivalence classes reduces the
EM's work sevenfold without changing the estimates.

In simulations, ProteoEM recovered the true molecular composition,
whereas approaches that reduced each affinity trace to a hard yes/no
call introduced substantial errors. Performance was robust to moderate,
uniform calibration error but was biased by informative missing data and
by proteoforms absent from the reference set. When observation yields
were known, ProteoEM also recovered source-sample composition from
observed molecular counts. ProteoEM is an open-source, reproducible
framework for quantitative analysis of single-molecule affinity
measurements and provides a basis for validation using experimental
molecule-level data.

\textbf{Keywords:} expectation-maximization; iterative affinity mapping;
single-molecule proteomics; protein inference; proteoforms; ambiguity;
affinity trace; weighted counting

\section{1. Introduction}\label{introduction}

Modern proteomics measures proteins at scale, providing broad
quantitative profiles from complex samples {[}1,2{]}. Yet proteomics has
not matched the throughput and molecular resolution of RNA sequencing.
This is because proteins have no equivalent of PCR amplification, their
abundances span a wide range, and each exists in many sequence,
processing, and modification states. Such variety makes it difficult to
determine which features coexist on one molecule. Mass spectrometry is
the mainstay of untargeted proteomics {[}1,2{]}, and its usual approach
is bottom-up, digesting each protein into peptides and measuring them.
Although it scales well, it breaks the linkage between parts of the same
molecule and leaves peptides that several proteins could explain
{[}3{]}. Top-down mass spectrometry instead measures the intact protein
and keeps that linkage, but it is harder to scale {[}4,5{]}. High-plex
affinity platforms take a different route and read DNA reporters:
SomaScan uses modified aptamers {[}6{]}, Olink and NULISA use antibody
pairs whose DNA tags are amplified and sequenced {[}7,8{]}, and Seer
compresses dynamic range with a nanoparticle protein corona before mass
spectrometry {[}9{]}. Each achieves high multiplexing within its design,
but its signal reports the abundance of a panel target or peptide and
stops short of the intact molecule's sequence and modification state.

Single-molecule approaches, including protein fluorosequencing, instead
aim to read individual molecules directly {[}10--12{]}. Iterative
single-molecule affinity mapping is one such approach {[}13--15{]}. It
infers protein identity and abundance from linked measurements on
individual molecules, without directly sequencing their amino acids.
Because each protein stays intact and is probed for several features,
the measurements preserve the within-molecule linkage that top-down mass
spectrometry provides, but at single-molecule resolution. The ordered
responses of one molecule across probe cycles form its affinity trace.
Figure 1 previews the assay, and Section 2.1 specifies the measurement
model.

An affinity trace is often compatible with more than one protein or
proteoform. A probe binds one feature, such as a short peptide motif or
a modification-specific segment, and many proteins and proteoforms carry
that feature, so a positive call narrows a trace to a set of candidates
rather than one. Probe binding is also imperfect. A probe can miss a
feature that is present or fire on one that is absent, so a trace is
never cleanly consistent or inconsistent with a candidate, and every
candidate keeps some likelihood. Because a yes/no call records only
presence or absence, the amount of that likelihood cannot be read from
the trace. It comes instead from an emission matrix \(Q\) calibrated
beforehand on known origins. When two candidates give the same pattern
of calls under every probe, no data can separate them, and a trace can
place them only as a group. RNA-seq quantification faces the same
ambiguity: a read can come from several transcripts, and
expectation-maximization (EM) splits each read's count among them
{[}16--20{]}. Supplementary Note S1 gives the full correspondence (Table
S1).

Given a set of these traces, the goal is to infer which proteins or
proteoforms are present and in what proportions. In a large-scale
single-molecule analysis of tau, this quantification resolved 130
distinct proteoform groups across model systems and human Alzheimer's
disease brain and revealed ordered, site-specific phosphorylation
{[}15{]}. PrISM scores how well each candidate protein explains a
molecule's binding and assigns the molecule to its single best-scoring
candidate before counting {[}13{]}. The tau IMaP EM estimates probe
binding rates jointly with abundance from the sample itself {[}15{]}.
Together these studies established weighted affinity evidence and
EM-based counting. Neither approach holds a pre-calibrated emission
matrix fixed and separate from the abundance step. Here we present
ProteoEM, an open-source Python package for weighted proteoform
abundance estimation from iterative affinity traces. It fixes \(Q\) and
keeps it separate from abundance estimation, retains each trace's
likelihood over all candidates instead of forcing a hard yes/no, reports
emission-equivalent candidates as a single group, and separates the
composition of the accepted molecules from that of the source sample.
Although this manuscript uses only a targeted panel, the same approach
applies to a proteome-wide reference, where only the candidate set is
larger.

We specify the observation model and evaluate ProteoEM in simulation, on
a panel similar to the tau IMaP assay with 12 probes over 36 cycles and
768 candidate proteoform states. We measure how well it identifies
proteoforms and recovers their abundance, and we compare it against
simpler baselines that force a hard yes/no for each candidate. We also
examine its behavior when the model's assumptions are violated. These
are simulation results; experimental validation remains future work.

\section{2. Methods}\label{methods}

\subsection{2.1 What ProteoEM estimates and
observes}\label{what-proteoem-estimates-and-observes}

In iterative affinity mapping, each molecule is immobilized at a
registered address on the chip and is probed over many cycles. In each
cycle a single affinity probe, such as an antibody or a multi-affinity
binding reagent, is applied, allowed to bind, imaged by fluorescence,
then removed before the next cycle. A probe binds one selected feature,
such as a short peptide motif of a few amino acids that recurs in many
proteins, or a modification- or isoform-specific segment that helps
define a proteoform. A bright spot at an address is a positive call and
its absence a negative call, and one image reads many addresses at once,
so a single cycle produces one column of calls across the molecules
(Fig. 1A, 1B). Different origins can share a feature, so one probe
cannot resolve them: PF-A and PF-C both carry the feature that probe
\(P_2\) queries, and \(P_2\) alone cannot tell them apart. The
registered molecules are not protein names, and several may be copies of
the same protein.

\begin{figure}
\centering
\includegraphics[width=1\linewidth,height=\textheight,keepaspectratio,alt={Reference features, physical cycles, and observed molecular traces. (A) The toy matrix E records ideal reference feature content; shared sequence segment A makes P\_2 alone ambiguous. (B) One physical P\_2 cycle gives one idealized call per registered teaching molecule, a single column of the observed matrix. (C) Applying P1--P5 once creates cycles c\_1--c\_5 of Y. Each M row follows one unknown-origin molecule. NA means that no usable positive or negative call was obtained and is not recoded as 0. (D) Reapplying P5 adds c\_6 without replacing c\_5=\textbackslash mathrm\{NA\}. Rows expand with registered molecules and columns with physical cycles. The matrix motivates likelihoods for individual traces and abundance estimation across accepted molecules.}]{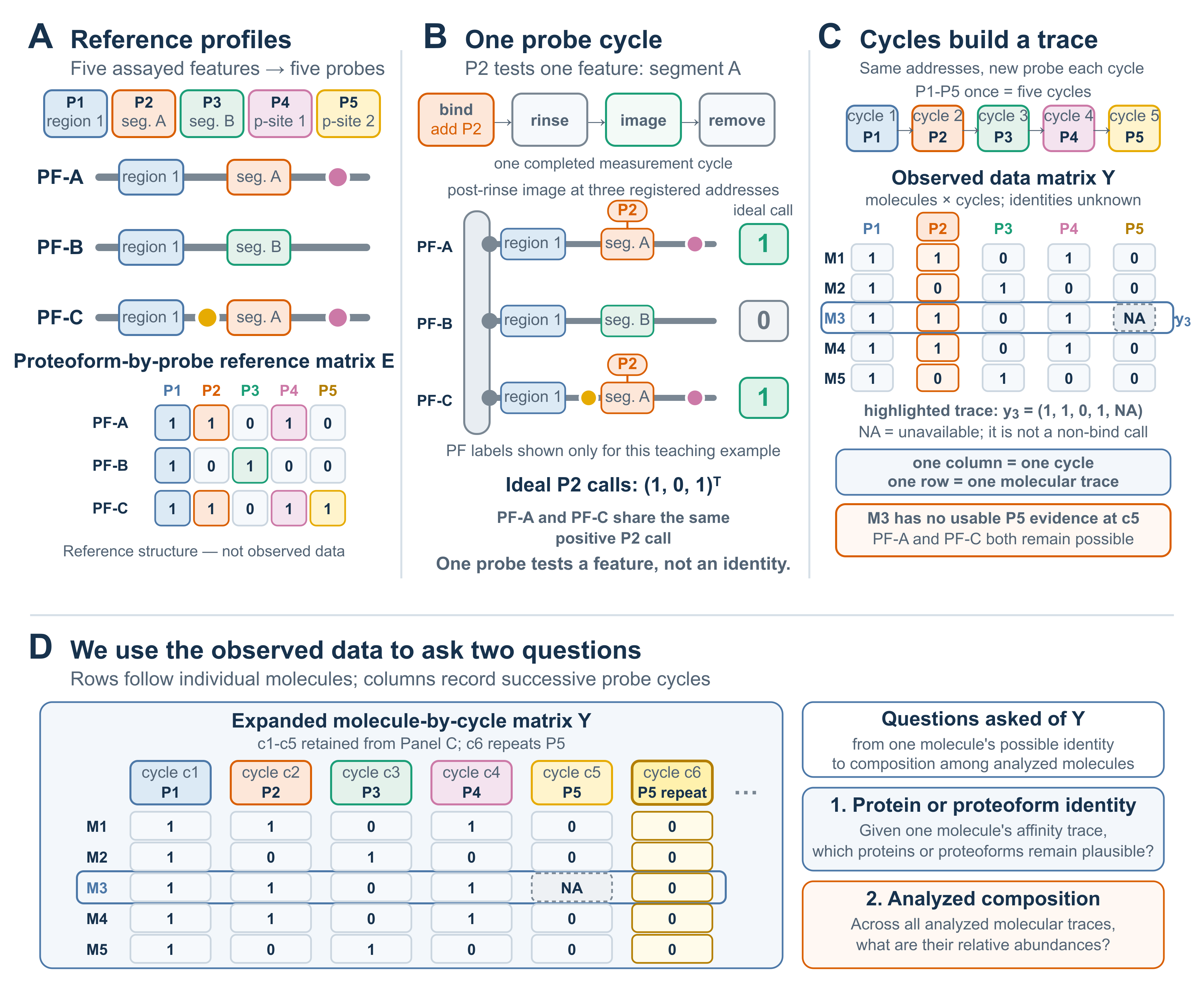}
\caption{Reference features, physical cycles, and observed molecular
traces. (A) The toy matrix \(E\) records ideal reference feature
content; shared sequence segment A makes \(P_2\) alone ambiguous. (B)
One physical \(P_2\) cycle gives one idealized call per registered
teaching molecule, a single column of the observed matrix. (C) Applying
P1--P5 once creates cycles \(c_1\)--\(c_5\) of \(Y\). Each M row follows
one unknown-origin molecule. NA means that no usable positive or
negative call was obtained and is not recoded as 0. (D) Reapplying P5
adds \(c_6\) without replacing \(c_5=\mathrm{NA}\). Rows expand with
registered molecules and columns with physical cycles. The matrix
motivates likelihoods for individual traces and abundance estimation
across accepted molecules.}
\end{figure}

We analyze the \(N\) accepted molecules, those that passed the assay's
quality filters. They come from a single sample, whether one donor's
plasma or plasma pooled from a patient cohort. Multiple samples are
quantified the same way, one fit each. Each molecule is read on its own:
it produces one trace across \(C\) physical cycles and enters the
estimate as a single count, and the assay uses \(J\) distinct logical
probes. Because each molecule is read individually, abundance is a count
of molecules, and each trace preserves the combination of features that
defines a proteoform. The reference set contains \(K\) proteins or
proteoforms that could have generated an observed trace; the model
assumes that every analyzed molecule came from one of them. The map
\(j(c)\in\{1,\ldots,J\}\) records which logical probe is used in
physical cycle \(c\). Reapplying a probe adds another cycle under the
same logical probe. The example in Figures 1 and 2 has \(N=5\), \(C=6\),
\(J=5\), and \(K=3\), with probe schedule \((P_1,P_2,P_3,P_4,P_5,P_5)\):
cycles \(c_5\) and \(c_6\) are separate measurements using the same
logical probe.

Let \(z_i\) be the unknown origin of accepted molecule \(i\). We model
these origins as

\[
z_i\mid\boldsymbol{\pi}
\overset{\mathrm{iid}}{\sim}
\operatorname{Categorical}(\boldsymbol{\pi}),
\qquad
\boldsymbol{\pi}=(\pi_1,\ldots,\pi_K),
\qquad
\pi_k\geq0,
\qquad
\sum_{k=1}^{K}\pi_k=1.
\]

The model treats the accepted molecules as independent draws from a
single composition \(\boldsymbol\pi\), where \(\pi_k\) is the proportion
of molecules arising from origin \(k\). Given the observed traces,
ProteoEM's EM (Section 2.4) returns a point estimate
\(\widehat{\boldsymbol\pi}\) of this relative abundance. What we
ultimately want is the source composition \(\boldsymbol\theta\): the
true proportion of each proteoform in the original sample. The two
differ because the assay does not accept every molecule equally: a
proteoform that is harder to recover, or more likely to fail the quality
filters, is underrepresented among the accepted molecules. Section 2.3
relates \(\boldsymbol\pi\) to \(\boldsymbol\theta\) through externally
calibrated effective observation yields.

\textbf{Observable proteoform groups.} Some proteoforms produce the same
distribution of accepted traces and cannot be separated by any probe in
the panel. We call such origins an \emph{observable proteoform group};
only their combined abundance is identifiable. The software finds groups
with identical emissions and reports their combined estimates
automatically. For example, two proteoforms that differ only at an
unmeasured site remain unresolved unless that difference also changes a
measured probe response. Trace equivalence does not imply equal
effective observation yields, and unequal yields do not make the
individual analyzed abundances identifiable.

For molecule \(i\) in physical cycle \(c\), let \(x_{ic}\in\{0,1\}\) be
the positive or negative call the probe would give for that cycle,
whether or not it can actually be read. The observed value is

\[
y_{ic}\in\{0,1,\mathrm{NA}\},
\]

where 1 is a positive call, 0 is a negative call, and NA means that no
usable call is available. An NA is not a negative call. When the call is
available, \(y_{ic}=x_{ic}\). Define

\[
o_{ic}=\mathbb{1}(y_{ic}\neq\mathrm{NA}),
\qquad
\mathcal O_i=\{c:o_{ic}=1\}.
\]

The main inputs are the \(N\times C\) trace matrix
\(\mathbf Y=(y_{ic})\) and the ordered probe schedule \(j(c)\) (Fig. 1C,
1D). Each row \(\mathbf y_i\) is one complete molecular trace. Repeated
probe applications occupy separate columns, and unavailable calls remain
distinct from observed negative calls.

\subsection{2.2 From calibrated probe responses to trace
likelihoods}\label{from-calibrated-probe-responses-to-trace-likelihoods}

To read a trace, we first need to know how the probes behave: how often
each probe lights up on each candidate identity. Probes are imperfect. A
probe can miss its target (a false negative) or bind off-target (a false
positive), so we summarize its behavior as a probability. For a molecule
of origin \(k\), \(q_{kj}\) is the chance that probe \(j\) gives a
positive call. These probabilities are typically measured experimentally
beforehand, by calibrating each probe on known reference proteins.
ProteoEM uses this measured \(Q\) as given; an empirically calibrated
\(Q\) can be supplied directly. Formally, for a physically recovered
molecule from origin \(k\), before any trace-dependent retention gate,

\[
q_{kj}=P(x_{ic}=1\mid z_i=k,\ j(c)=j,\ R_i=1).
\]

So \(q_{kj}=0.90\) means that repeated applications of that probe to
molecules from origin \(k\) are called positive about 90\% of the time.
The \(K\times J\) matrix \(\mathbf Q=(q_{kj})\) is then frozen during
the abundance EM and never learned from the unknown mixture (Fig. 2B).
This is what \emph{fixed emissions} means. The tau IMaP EM instead
estimates binding rates jointly with abundance from the sample {[}15{]}.
Fixing \(Q\) keeps calibration separate from abundance estimation and
auditable, and it relies on that calibration being correct. In our
model, every physical cycle using logical probe \(j\) uses the same
probability \(q_{kj}\), but each application remains a separate
observation; if calibration shows that a repeat, cycle, or batch behaves
differently, a pre-calibrated cycle- or batch-specific probability may
be supplied instead.

Calibrating every cell of \(Q\) directly can be impractical when there
are many origins and probes. A validated structured model can then fill
the cells in by sharing information across them. The simplest structure
starts from a binary feature matrix \(\mathbf E=(E_{kj})\), where
\(E_{kj}=1\) means origin \(k\) is expected to carry the feature that
probe \(j\) queries. Give each probe two rates: an on-target
positive-call rate \(\alpha_j\) and an off-target rate \(\beta_j\). Then

\[
q_{kj}=E_{kj}\alpha_j+(1-E_{kj})\beta_j.
\tag{1}
\]

For example, if \(\alpha_j=0.90\) and \(\beta_j=0.10\), a
feature-positive origin receives \(q_{kj}=0.90\) and a feature-negative
origin receives \(q_{kj}=0.10\). Equation 1 is a simplification that
does not substitute for validating the emission model: a binary feature
indicator does not automatically represent site number, accessibility,
position, sequence context, steric effects, cycle, or batch.
Constructing \(Q\) from adequate known-origin standards, and the
conditions under which a structured model is admissible, are described
in Supplementary Methods S2.1. Regardless of construction route, \(Q\)
remains fixed during mixture analysis.

\textbf{From one trace to one likelihood row.} For each candidate
origin, we want the likelihood of the trace, the probability that this
origin would produce the calls we observed. Our model computes it under
two assumptions: every independently registered trace is kept, and, once
the origin and \(Q\) are fixed, the cycle calls are independent of one
another. Independence lets us multiply the per-cycle probabilities. Each
observed cycle contributes one factor: the probe's positive-call rate
\(q\) if the call was positive, or \(1-q\) if it was negative (Fig. 2C).
Multiplying these across the observed cycles gives \(L_{ik}\), the
likelihood of trace \(i\) under origin \(k\):

\[
L_{ik}
=P(\mathbf y_{i,\mathcal O_i}\mid z_i=k,\mathbf Q,j(\cdot))
=\prod_{c\in\mathcal O_i}
q_{k,j(c)}^{y_{ic}}
\left(1-q_{k,j(c)}\right)^{1-y_{ic}}.
\tag{2}
\]

Each observed physical cycle contributes separately, including repeated
applications of the same logical probe. For stable computation, ProteoEM
adds log-likelihood contributions instead of multiplying many small
numbers:

\[
\log L_{ik}
=\sum_{c\in\mathcal O_i}
\left[
y_{ic}\log q_{k,j(c)}
+(1-y_{ic})\log\left(1-q_{k,j(c)}\right)
\right],
\tag{3}
\]

with the convention \(0\log0=0\); an observed result assigned
probability zero has log likelihood \(-\infty\). The row \(L_{i\cdot}\)
compares how well the possible origins explain one trace; it is not yet
the probability that the molecule came from each origin. The abundance
mixture enters that calculation in Section 2.4.

\textbf{Missing cycle calls.} An NA means that no usable call was
obtained. It can be omitted only when missingness, after accounting for
recorded technical factors, is unrelated to both the unseen call and the
molecular origin: \emph{ignorable missingness}, formally
\(o_{ic}\perp(x_{ic},z_i)\mid\mathbf w_{ic}\) for relevant covariates
\(\mathbf w_{ic}\). The unseen binary call is then summed over:

\[
\sum_{a\in\{0,1\}}
q_{k,j(c)}^{a}
\left(1-q_{k,j(c)}\right)^{1-a}
=q_{k,j(c)}+\left(1-q_{k,j(c)}\right)
=1.
\tag{4}
\]

A factor of one changes no trace likelihood, so an unavailable cycle may
be omitted from Equations 2 and 3; displaying it as \(\times 1\) simply
makes that neutral contribution explicit, while other observed
applications of the same probe remain informative. When an NA depends on
the hidden call or the origin, for example when a bright spot saturates
and suppresses positive calls, it cannot be ignored, and the required
explicit observed-data model is given in Supplementary Methods S2.2.

\textbf{Worked example: the \(M_3\) trace.} We follow the third trace
from Figure 1, \(\mathbf y_3=(1,1,0,1,\mathrm{NA},0)\), through the
schedule \((P_1,P_2,P_3,P_4,P_5,P_5)\) (Fig. 2A). For PF-A, the positive
calls in \(c_1\), \(c_2\), and \(c_4\) each contribute 0.90, the
negative calls in \(c_3\) and \(c_6\) each contribute \(1-0.10=0.90\),
and the unavailable \(c_5\) call contributes one, giving
\(L_{3,A}=0.90^5=0.59049\). PF-C shares PF-A's response probabilities
for \(P_1\)--\(P_4\) but has positive-call probability 0.90 rather than
0.10 for \(P_5\); the missing \(c_5\) call cannot distinguish them,
whereas the observed negative \(c_6\) call contributes 0.10 for PF-C and
0.90 for PF-A. Applying the same calculation to all three origins gives

\[
L_{3\cdot}=(0.59049,0.00081,0.06561).
\]

These values need not sum to one; they compare how well each origin
explains \(M_3\) and are not yet probabilities of molecular origin.
Because this teaching example retains every registered trace, the
displayed products are also accepted-trace likelihoods; Section 2.3
gives the required conditioning when trace responses determine
retention.

\begin{figure}
\centering
\includegraphics[width=1\linewidth,height=\textheight,keepaspectratio,alt={From known-origin calibration and one molecular trace to a likelihood row. (A) The six-cycle M\_3 trace is paired with its cycle-to-probe map; c\_5 and c\_6 are separate P\_5 observations. (B) Synthetic known-origin control counts illustrate how positive-call rates can populate a fixed Q; a validated structured model is an alternative when direct cell-wise calibration is impractical. Q is calibrated before mixture analysis, frozen during abundance EM, and not learned from the unknown mixture. (C) Each origin uses the complete factor calculation: a positive call selects q, a negative call selects 1-q, and, under ignorable missingness, an NA is neutral (\textbackslash times1). (D) The products form L\_\{3\textbackslash cdot\}, the evidence across candidate origins. The population mixture enters in Figure 3. All-trace retention is assumed here; gated data require the joint-likelihood conditioning in Section 2.3. All displayed control counts and 0.90/0.10 values are synthetic teaching values.}]{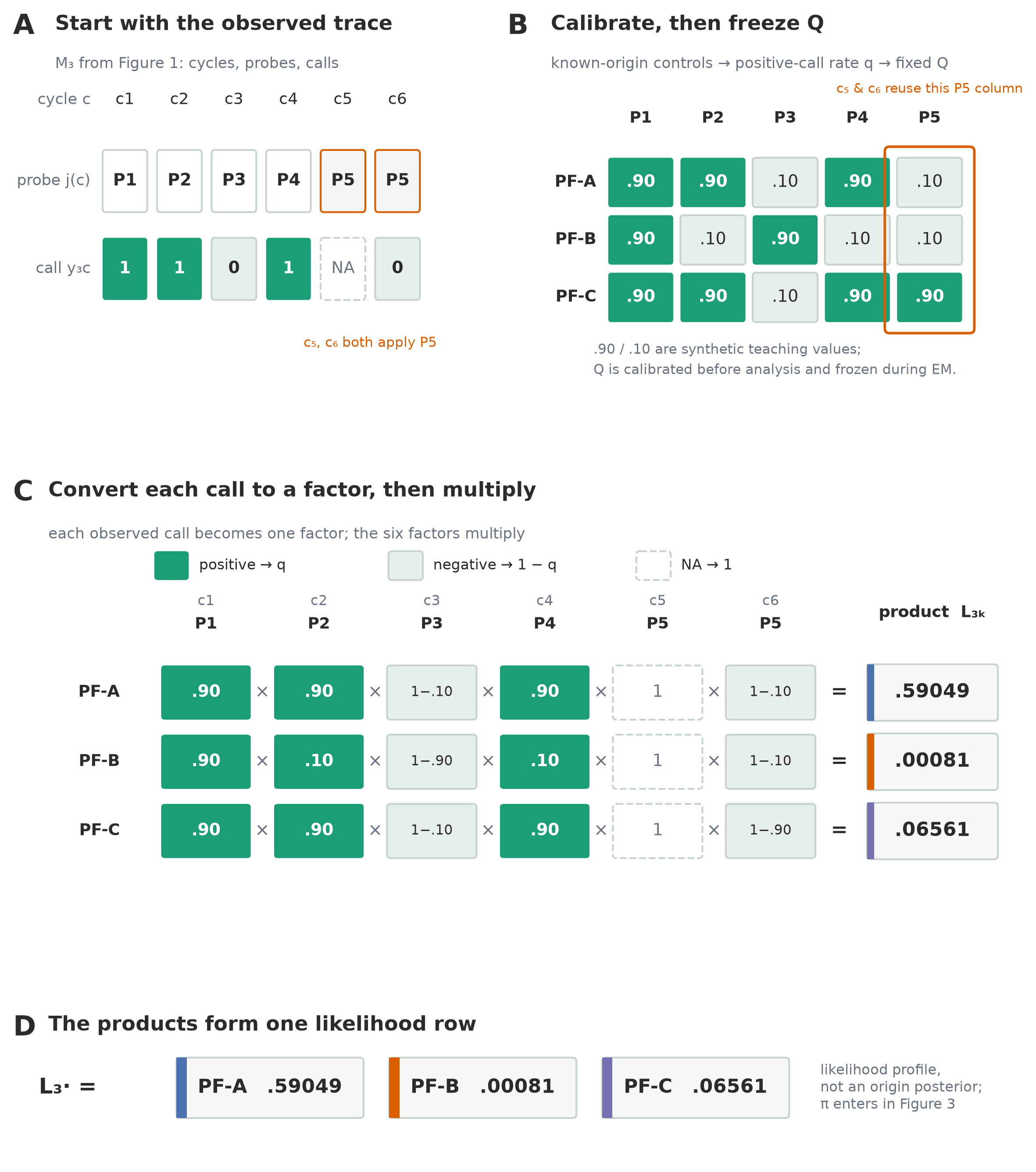}
\caption{From known-origin calibration and one molecular trace to a
likelihood row. (A) The six-cycle \(M_3\) trace is paired with its
cycle-to-probe map; \(c_5\) and \(c_6\) are separate \(P_5\)
observations. (B) Synthetic known-origin control counts illustrate how
positive-call rates can populate a fixed \(Q\); a validated structured
model is an alternative when direct cell-wise calibration is
impractical. \(Q\) is calibrated before mixture analysis, frozen during
abundance EM, and not learned from the unknown mixture. (C) Each origin
uses the complete factor calculation: a positive call selects \(q\), a
negative call selects \(1-q\), and, under ignorable missingness, an NA
is neutral (\(\times1\)). (D) The products form \(L_{3\cdot}\), the
evidence across candidate origins. The population mixture enters in
Figure 3. All-trace retention is assumed here; gated data require the
joint-likelihood conditioning in Section 2.3. All displayed control
counts and 0.90/0.10 values are synthetic teaching values.}
\end{figure}

\clearpage

\subsection{2.3 Effective observation yield and accepted-trace
composition}\label{effective-observation-yield-and-accepted-trace-composition}

A molecule's trace is counted only if two things happen: the molecule is
recovered and the trace passes the retention rule. Some molecules clear
both more readily than others. The \textbf{effective observation yield}
is the expected number of accepted traces one source molecule produces
under the assay and its retention rule. For example, under a rule that
keeps a trace only when at least one call is positive, a proteoform with
many bindable probe opportunities has more chances to register a
positive and is retained more often than one with few; equal source
amounts of the two then produce unequal accepted counts. Because an
intact molecule produces at most one trace, this yield is a probability
between zero and one. It plays the same bookkeeping role as effective
transcript length in RNA-seq. There, read counts are divided by a
transcript's effective length to recover abundance; here, accepted
counts are divided by the effective observation yield.

Let \(R_i=1\) mean that source molecule \(i\) is physically recovered,
registered, and available for measurement, with
\(r_k=P(R_i=1\mid z_i=k)\). For a recovered molecule, let
\(f_k(\mathbf y)=P(\mathbf Y=\mathbf y\mid z_i=k,R_i=1)\) be the
\emph{potential-trace distribution} before any whole-trace gate, let
\(g(\mathbf y)\in[0,1]\) be the probability that trace \(\mathbf y\) is
retained, and let \(A_i=1\) denote retention. Visibility and total
effective observation yield are

\[
v_k
=P(A_i=1\mid z_i=k,R_i=1)
=\sum_{\mathbf y}g(\mathbf y)f_k(\mathbf y),
\qquad
e_k=r_kv_k.
\]

Thus \(r_k\) describes physical recovery, \(v_k\) describes selection
based on the trace, and \(e_k\) describes the complete
source-to-accepted-trace path. If \(\theta_k\) is the source
composition, then the accepted-trace composition is
\(\pi_k=\theta_ke_k/\sum_h\theta_he_h\), and for origins with \(v_k>0\)
the likelihood used by EM must match that accepted population:

\[
L^{\mathrm{acc}}_{ik}
=P(\mathbf Y=\mathbf y_i\mid z_i=k,R_i=1,A_i=1)
=\frac{g(\mathbf y_i)f_k(\mathbf y_i)}{v_k}.
\]

If every independently registered trace is retained, including
all-negative traces, then \(g\equiv1\), \(v_k=1\), and the likelihood in
Equation 2 is already \(L^{\mathrm{acc}}_{ik}=f_k(\mathbf y_i)\). In
that design, a protein with 12 bindable probe opportunities still
produces one trace, just as a protein with 3 opportunities does; the
difference changes \(Q\), the trace pattern, and decodability, but not
trace yield. If instead at least one recorded positive call is required,
probe opportunity changes trace yield. With fully observed independent
cycles, \(v_k=1-\prod_c(1-q_{k,j(c)})\): for equal per-opportunity
\(q=0.20\) and equal recovery, 12 opportunities give
\(v_A=1-0.8^{12}=0.931\) and 3 give \(v_B=1-0.8^3=0.488\), so equal
source amounts appear as approximately 65.6\% A and 34.4\% B among
retained traces.

When every \(e_k>0\) is externally calibrated, relative source
composition is recovered by \(\theta_k=(\pi_k/e_k)/\sum_h(\pi_h/e_h)\);
only relative yields are needed. If \(e_k=0\), that source is
unobservable. Acceptance must not mean ``successfully identified'': all
traces passing prespecified physical and technical criteria remain in
the mixture fit, even when their posterior assignment probabilities are
diffuse. Retention conditioning by \(v_k\) and inverse-yield correction
by \(e_k\) do different jobs and neither should be applied twice. A
trace-dependent gate can also make calls that were independent become
dependent, so the trace likelihood must then be conditioned on
acceptance. Both corrections, why they are not interchangeable, and the
conditions a supplied likelihood must meet are given in Supplementary
Methods S2.3.

\subsection{2.4 Estimating analyzed composition with
EM}\label{estimating-analyzed-composition-with-em}

ProteoEM estimates abundance by repeating two steps until the estimate
stops changing. The E-step splits each molecule's single count across
the origins that could have produced its trace, in proportion to how
well each origin explains it. The M-step adds up those fractional counts
over all molecules to get an updated composition, which feeds back into
the next E-step (Fig. 3D). Here \emph{abundance} means relative
composition among the accepted, analyzed molecules.

With independent molecules and fixed trace likelihoods (\(L_{ik}\geq0\),
\(\sum_kL_{ik}>0\) for every retained row), the likelihood and
log-likelihood of the analyzed composition \(\boldsymbol\pi\) are

\[
\mathcal L(\boldsymbol\pi;\mathbf L)
=\prod_{i=1}^{N}\sum_{k=1}^{K}\pi_kL_{ik},
\tag{5}
\]

\[
\ell(\boldsymbol\pi)
=\sum_{i=1}^{N}
\log\left(\sum_{k=1}^{K}\pi_kL_{ik}\right).
\tag{6}
\]

For molecule \(i\), the sum \(\sum_k\pi_kL_{ik}\) combines all possible
origins: \(L_{ik}\) measures how well origin \(k\) explains the trace,
and \(\pi_k\) supplies its current abundance weight.

\textbf{E-step.} Given the current estimate \(\boldsymbol\pi^{(s)}\),
the posterior assignment probability (the \emph{responsibility})

\[
\gamma_{ik}^{(s)}
=P(z_i=k\mid\mathbf y_i,A_i=1,\boldsymbol\pi^{(s)})
=\frac{\pi_k^{(s)}L_{ik}}
{\sum_{\ell=1}^{K}\pi_\ell^{(s)}L_{i\ell}}
\tag{7}
\]

is the model-based probability that accepted molecule \(i\) has origin
\(k\), and also the fraction of that molecule's one count assigned to
origin \(k\); \(\sum_k\gamma_{ik}^{(s)}=1\). Figure 3A uses the
illustrative binary-limit row \(L_{i\cdot}=(1,0,1)\): with uniform
initialization PF-A and PF-C each receive \(0.50\); at a later iteration
with \(\boldsymbol\pi^{(s)}=(.60,.30,.10)\), Equation 7 gives
\(\gamma_{i\cdot}^{(s)}\approx(.86,0,.14)\) (Fig. 3B). The ambiguous
trace has borrowed information from the rest of the dataset through the
shared abundance estimate, even though its likelihood row did not
change.

\textbf{M-step.} The expected count for origin \(k\) is the sum of its
posterior assignment probabilities across all molecules; dividing by
\(N\) gives the next estimate:

\[
\pi_k^{(s+1)}
=\frac{1}{N}\sum_{i=1}^{N}\gamma_{ik}^{(s)}.
\tag{8}
\]

For example, if \(N=100\) traces give expected counts \((64,27,9)\), the
updated composition is \((.64,.27,.09)\) (Fig. 3C). At an exact EM fixed
point the posterior-weighted count satisfies

\[
\widehat n_k=\sum_{i=1}^{N}\widehat\gamma_{ik}
=N\widehat\pi_k,
\tag{9}
\]

and numerical fits satisfy Equation 9 up to the convergence tolerance.
For fixed \(\mathbf L\), Equation 6 is concave on the probability
simplex, so it has no inferior local maxima, although the maximum can be
non-unique when component distributions are identical or linearly
dependent. Numerical implementation, convergence criteria, and an
independent projected-gradient check are given in Supplementary Methods
S2.6.

\begin{figure}
\centering
\includegraphics[width=1\linewidth,height=\textheight,keepaspectratio,alt={How all accepted traces update analyzed composition. (A) A new binary-limit teaching example, separate from the real-valued M\_3 likelihood row in Figure 2, supplies one illustrative likelihood profile. (B) Equal splitting occurs initially here only because the supported likelihoods and starting analyzed-composition weights are equal. After all accepted traces update the shared composition, the same row receives an unequal analyzed-composition-weighted split. (C) Pooling the posterior-assignment-probability vectors from all N accepted traces gives expected counts; division by N gives the next analyzed composition. The illustrative N=100 example gives \textbackslash boldsymbol n=(64,27,9) and \textbackslash boldsymbol\textbackslash pi\^{}\{(s+1)\}=(.64,.27,.09). (D) That updated composition returns to the repeated E-step.}]{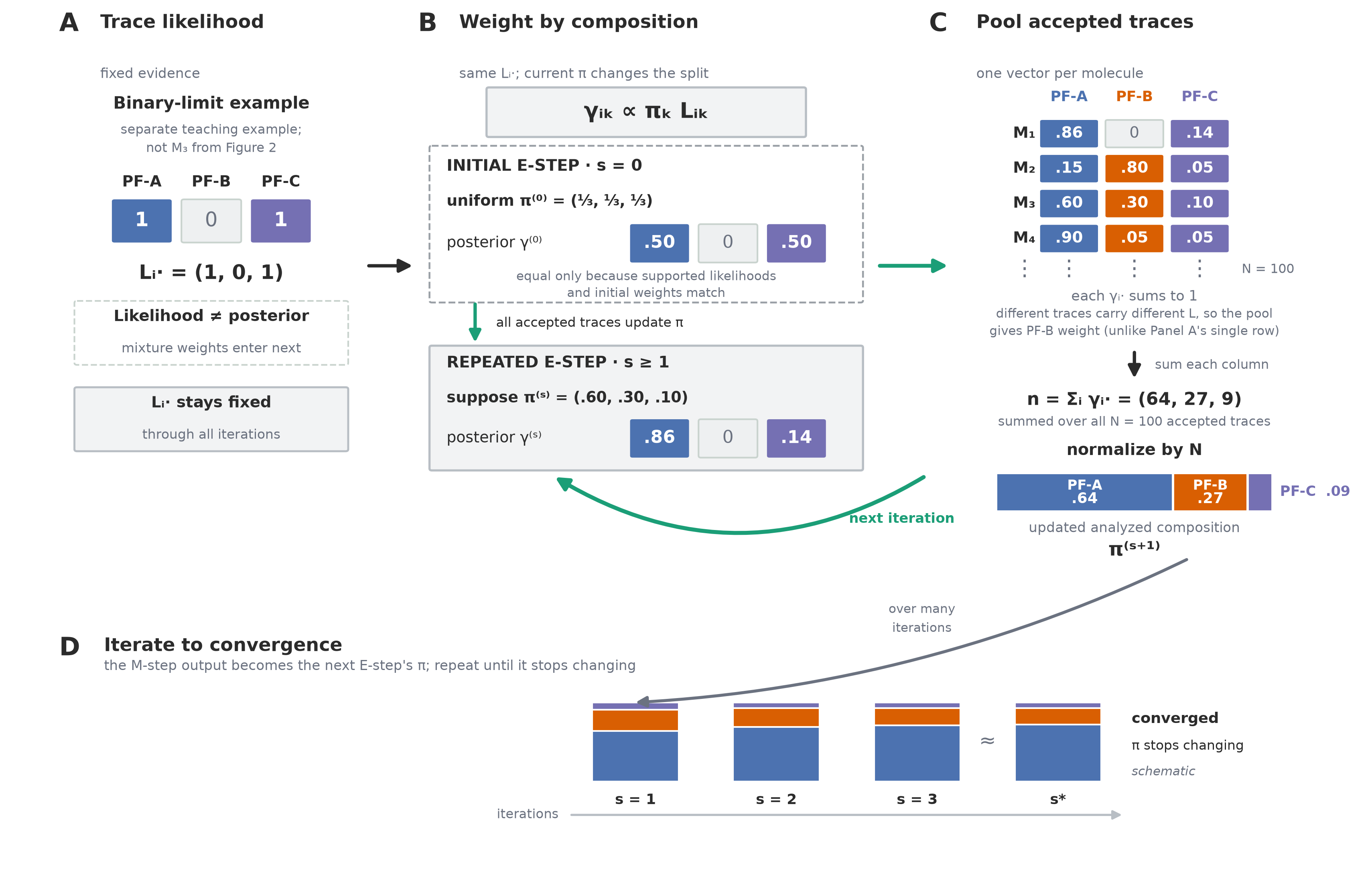}
\caption{How all accepted traces update analyzed composition. (A) A new
binary-limit teaching example, separate from the real-valued \(M_3\)
likelihood row in Figure 2, supplies one illustrative likelihood
profile. (B) Equal splitting occurs initially here only because the
supported likelihoods and starting analyzed-composition weights are
equal. After all accepted traces update the shared composition, the same
row receives an unequal analyzed-composition-weighted split. (C) Pooling
the posterior-assignment-probability vectors from all \(N\) accepted
traces gives expected counts; division by \(N\) gives the next analyzed
composition. The illustrative \(N=100\) example gives
\(\boldsymbol n=(64,27,9)\) and
\(\boldsymbol\pi^{(s+1)}=(.64,.27,.09)\). (D) That updated composition
returns to the repeated E-step.}
\end{figure}

\clearpage

\textbf{Resolution-aware reporting.} Origins \(k\) and \(\ell\) belong
to the same observable proteoform group when they generate the same
accepted-trace distribution:

\[
P(\mathbf y\mid z=k,A=1)=P(\mathbf y\mid z=\ell,A=1)
\quad\text{for every observable trace }\mathbf y.
\tag{10}
\]

The mixture likelihood then depends on \(\pi_k\) and \(\pi_\ell\) only
through \(\pi_k+\pi_\ell\), so their internal split is unidentified even
after EM converges; unequal effective observation yields do not make
emission-equivalent origins distinguishable. ProteoEM detects exact
duplicate scheduled-emission rows and returns observable groups together
with group-level weights, expected counts, and posterior assignment
probabilities (Fig. S1). Adding candidates to the reference introduces
more hypotheses but does not improve resolution unless the measurements
can distinguish them. Near-identifiability diagnostics and the treatment
of trace gates in Equation 10 are given in Supplementary Methods S2.7.

\textbf{Exact compression and comparison methods.} Molecules \(i\) and
\(i'\) whose likelihood rows are proportional across every origin,

\[
L_{ik}=a_{ii'}L_{i'k}\ \text{for all }k,\qquad a_{ii'}>0,
\tag{11}
\]

form a \emph{trace-likelihood class}: the positive row-wide scale
cancels from Equation 7, so they share posterior assignment
probabilities at every iteration. EM then processes the \(T\le N\)
distinct class profiles, reducing one dense iteration from \(O(NK)\) to
\(O(TK)\) while preserving the fitted composition exactly. This row
grouping is distinct from the observable proteoform groups of Equation
10, which combine columns; the exact class decomposition is derived in
Supplementary Methods S2.4. As deliberately simpler baselines, the
benchmark also runs binary-profile EM, which rounds the likelihoods to
hard 0/1 compatibility, and top-likelihood counting, which assigns each
molecule to its maximum-likelihood origin; both reduce each trace to a
hard yes/no and are defined in Supplementary Methods S2.5.

\section{3. Results}\label{results}

We apply ProteoEM to three tasks. First, we identify which proteoforms
are present and estimate their abundances among the analyzed molecules:
ProteoEM scores each molecule against every candidate proteoform and
uses EM to turn those scores into a per-molecule probability of each
origin and a shared abundance estimate, so identification and
quantification come from the same fit. Second, we check that the
estimate stays predictable when the modeling assumptions break. Third,
we correct for observation yield so the estimate reflects the original
sample and not only the molecules that produced usable traces.

\subsection{3.1 The simulated benchmark}\label{the-simulated-benchmark}

Every result in this section uses simulated data, because experimental
iterative-affinity datasets are not publicly available. Simulation has a
clear advantage: the true composition and probe behavior are known, so
we can measure how close an estimate comes to the truth and turn each
modeling assumption on or off by itself. It has a limitation: the
emission matrix \(Q\) is idealized and never measured from real probe
binding, and the data contain no calibration error, recovery variation,
or correlated noise unless we add them. These results therefore test the
estimator and its bookkeeping. They do not show how the assay would
perform.

Each dataset is drawn from the same generative model, laid out below.
Figure 4 follows the whole process, from the true composition to the
traces ProteoEM sees. The data are generated in two steps: first the
abundances set how many molecules come from each proteoform, then the
probe model sets the calls each molecule produces.

The reference set holds 768 candidate proteoform states: six tau isoform
backbones crossed with the \(2^7=128\) settings of seven binary
modification sites. Each simulated sample has 32 of the 768 proteoforms
present and the other 736 exactly zero. To keep every isoform
represented, one present proteoform is drawn from each of the six
isoforms and 26 more without replacement from the rest of the panel,
which is 32 in all. Their abundances come from a
\(\operatorname{Dirichlet}(0.4,\ldots,0.4)\) draw. The concentration of
\(0.4\) is below one, so each draw is uneven: a few present states carry
most of the abundance and many are rare (Figure 4A, 4B).

We then draw \(N\) molecules to turn these abundances into actual
molecule counts. Each molecule is given a true origin at random, with
each proteoform's chance equal to its abundance, so a common proteoform
supplies many molecules and a rare one few. That origin fixes the
molecule's feature content, its row of the reference matrix, and is then
discarded, never entering the data.

Each molecule is read by 12 logical probes applied three times, for 36
cycles. Every cycle yields one Bernoulli call: the probe fires with
probability \(\alpha_j\) if the origin carries the queried feature and
\(\beta_j\) if it does not (Figure 4C). Because the call is a coin flip,
a present feature sometimes reads negative and an absent one sometimes
reads positive, so repeating each probe lets the noise average out.
Finally, 2\% of calls are dropped at random and recorded as NA.

The result is one trace of 36 calls per molecule (Figure 4D). The
origins are gone, and a trace is a noisy pattern of positive, negative,
and missing calls that many candidates could have produced. A
proteoform's abundance survives only as how often its trace pattern
recurs across the molecules, so recovering the composition is a mixture
problem: ProteoEM weighs each trace against every candidate with the
known \(Q\) and pools those weights into an abundance estimate (Figure
4E).

In compact form, the model draws a composition, then for each molecule
its origin and its calls:

\[
\begin{aligned}
\boldsymbol\theta &\sim \operatorname{Dirichlet}(0.4,\ldots,0.4) && \text{on the present states, } \theta_k=0 \text{ otherwise},\\
z_i &\sim \operatorname{Categorical}(\boldsymbol\theta), && i=1,\ldots,N,\\
x_{ic} &\sim \operatorname{Bernoulli}\!\left(q_{z_i,\,j(c)}\right), && c=1,\ldots,C,\\
y_{ic} &= x_{ic} \text{ with probability } 1-p, \text{ otherwise } \mathrm{NA}.
\end{aligned}
\tag{12}
\]

Here \(z_i\) is the origin of molecule \(i\), one of the \(K\) candidate
proteoforms. Its features are that candidate's row of \(\mathbf E\),
which sets the call probabilities \(q_{z_i,j(c)}\) through Equation 1,
and \(p=0.02\) is the missingness rate. Only the composition, the
origins, and the calls are drawn at random: a molecule's features follow
deterministically from its origin. The origins \(z_i\) are never
observed, so ProteoEM estimates \(\boldsymbol\theta\) from the trace
matrix \(\mathbf Y\) alone by fitting this mixture (Section 2.4).

\begin{figure}
\centering
\includegraphics[width=1\linewidth,height=\textheight,keepaspectratio,alt={The generative model, from true composition to recovered abundance. Every dataset is drawn from the model, so the truth is known and recovery can be measured. (A) A Dirichlet(0.4) draw on a three-state simplex: a concentration below one pushes draws toward the corners and edges, where one or two states dominate, and away from the center, where all are equal, so each draw is skewed. (B) The true abundances \textbackslash boldsymbol\textbackslash theta for one seed: 32 present proteoforms with a heavy tail, and 736 held at exactly zero. (C) How one molecule reads. Its origin is drawn with probability proportional to \textbackslash boldsymbol\textbackslash theta and fixes the top row: the molecule's true features, one row of the reference matrix E (dark = feature present, pale = absent), which is the hidden truth and never enters the data. The bottom row is the resulting calls (green = positive, pale = negative): each present feature reads positive with probability \textbackslash alpha and each absent feature with probability \textbackslash beta, so the two rows agree except where noise flips a call. Repeating the 12 probes three times and dropping 2\% of calls as NA gives the molecule's 36-cycle trace. The rates \textbackslash alpha=0.9 and \textbackslash beta=0.1 are illustrative teaching values, as in Figure 2; the benchmark uses per-probe calibrated rates (Supplementary Methods S3.5). (D) A slice of the resulting call matrix, molecules by 36 cycles (12 probes applied in three rounds). (E) The weighted-EM estimate \textbackslash widehat\{\textbackslash boldsymbol\textbackslash pi\} laid over the truth \textbackslash boldsymbol\textbackslash theta; it tracks the profile to a total-variation error of 0.013. Panels B, D, and E use real values from one benchmark run (N=20\{,\}000, seed 7); panels A and C are illustrative. All values are simulated.}]{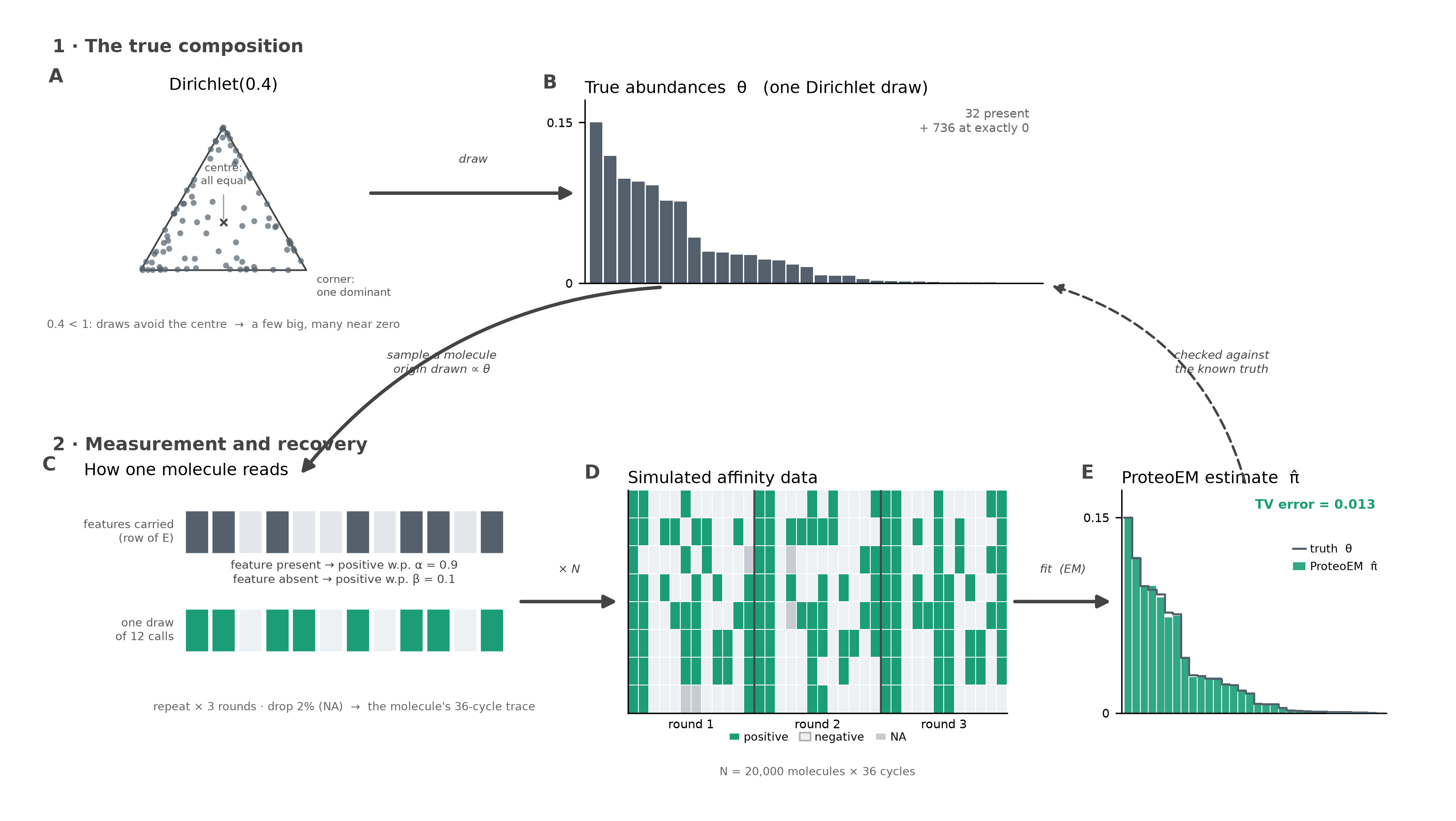}
\caption{\textbf{The generative model, from true composition to
recovered abundance.} Every dataset is drawn from the model, so the
truth is known and recovery can be measured. (A) A Dirichlet(0.4) draw
on a three-state simplex: a concentration below one pushes draws toward
the corners and edges, where one or two states dominate, and away from
the center, where all are equal, so each draw is skewed. (B) The true
abundances \(\boldsymbol\theta\) for one seed: 32 present proteoforms
with a heavy tail, and 736 held at exactly zero. (C) How one molecule
reads. Its origin is drawn with probability proportional to
\(\boldsymbol\theta\) and fixes the top row: the molecule's true
features, one row of the reference matrix \(E\) (dark = feature present,
pale = absent), which is the hidden truth and never enters the data. The
bottom row is the resulting calls (green = positive, pale = negative):
each present feature reads positive with probability \(\alpha\) and each
absent feature with probability \(\beta\), so the two rows agree except
where noise flips a call. Repeating the 12 probes three times and
dropping 2\% of calls as NA gives the molecule's 36-cycle trace. The
rates \(\alpha=0.9\) and \(\beta=0.1\) are illustrative teaching values,
as in Figure 2; the benchmark uses per-probe calibrated rates
(Supplementary Methods S3.5). (D) A slice of the resulting call matrix,
molecules by 36 cycles (12 probes applied in three rounds). (E) The
weighted-EM estimate \(\widehat{\boldsymbol\pi}\) laid over the truth
\(\boldsymbol\theta\); it tracks the profile to a total-variation error
of 0.013. Panels B, D, and E use real values from one benchmark run
(\(N=20{,}000\), seed 7); panels A and C are illustrative. All values
are simulated.}
\end{figure}

Correctly specified runs give the EM the same \(Q\) that generated the
data; Section 3.3 perturbs or degrades \(Q\) to test misspecification.

A second benchmark, with two origins, jointly varies physical recovery
and trace gating to test the source-composition correction of Section
3.4. The tau-inspired panel reproduces the size and broad logical
structure of the reported tau panel; it is not Nautilus data or
calibration. Supplementary Methods S3 gives the complete protocol: the
candidate panel, abundance and probe-response generation, the paired
model-violation scenario grid, comparison methods, and evaluation
metrics. The software package is summarized in Supplementary Software.

\subsection{3.2 Identifying proteoforms and estimating their
abundance}\label{identifying-proteoforms-and-estimating-their-abundance}

\textbf{Identifying each molecule's proteoform.} Both identifying each
molecule's proteoform and estimating each proteoform's abundance come
from one EM fit over all the accepted traces. For each accepted trace,
it points to the proteoform most likely to have produced it. ProteoEM
scores the trace against every candidate and, as calls accumulate over
repeated probing, concentrates the per-molecule posterior on the origin
that best explains them. Figure 5 follows one molecule: under the full
12-probe panel its posterior concentrates on the true 2N4R proteoform
over three rounds of probing, pulling away from its 2N3R twin. A single
probe carries that separation. Switch off the one probe that tells 3R
from 4R and the two proteoforms become indistinguishable: the posterior
splits evenly between them, while the total for the pair ProteoEM can
observe stays identified. Across the correctly specified baseline
(twenty-seed set, 10,000 accepted traces each), ProteoEM found which
proteoforms were present with sensitivity \(0.978\pm0.027\) and a
false-discovery rate of \(0.010\pm0.018\), put the true origin at the
top of the list for \(0.990\pm0.003\) of molecules, and within the top
five for essentially all of them. Its confidence was well calibrated:
expected calibration error 0.0011, Brier score 0.014.

\begin{figure}
\centering
\includegraphics[width=1\linewidth,height=\textheight,keepaspectratio,alt={Evidence accumulates over repeated probing to identify one molecule's proteoform. Flat-prior per-molecule posterior probability of each candidate origin, updated as calls accumulate over 36 cycles (12 logical probes applied in three rounds), for one simulated 2N4R molecule. (A) With the full panel the posterior concentrates on the true 2N4R proteoform (0.99), separating it from its 2N3R twin; the three cycles where the 3R/4R discriminator fires are marked with filled triangles. (B) With that discriminator removed (open triangles), 2N4R and 2N3R are emission-equivalent and the posterior splits evenly (0.50 each), while their observable group total stays identified (0.99). Faint grey traces are the other candidates. The staircase reflects noise-averaging across repeated applications of the same probes; the final assignment does not depend on probe order. All values are simulated.}]{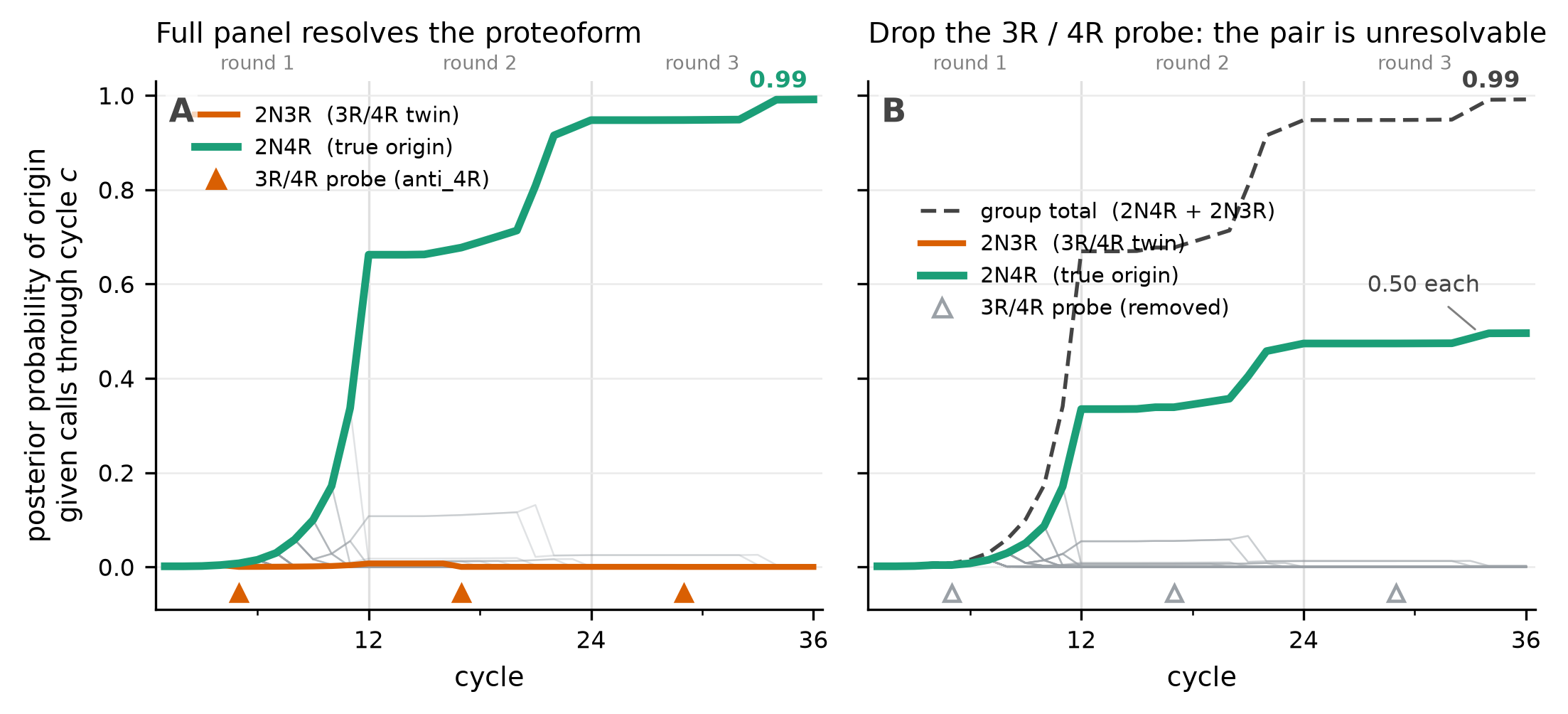}
\caption{\textbf{Evidence accumulates over repeated probing to identify
one molecule's proteoform.} Flat-prior per-molecule posterior
probability of each candidate origin, updated as calls accumulate over
36 cycles (12 logical probes applied in three rounds), for one simulated
2N4R molecule. (A) With the full panel the posterior concentrates on the
true 2N4R proteoform (0.99), separating it from its 2N3R twin; the three
cycles where the 3R/4R discriminator fires are marked with filled
triangles. (B) With that discriminator removed (open triangles), 2N4R
and 2N3R are emission-equivalent and the posterior splits evenly (0.50
each), while their observable group total stays identified (0.99). Faint
grey traces are the other candidates. The staircase reflects
noise-averaging across repeated applications of the same probes; the
final assignment does not depend on probe order. All values are
simulated.}
\end{figure}

\textbf{Convergence and implementation checks.} An abundance estimate is
only trustworthy if the code reaches the fit it claims to, so we checked
the implementation two ways. First, the grouping shortcuts that make
ProteoEM fast must not change the result: fitting one likelihood row per
molecule, grouping identical traces, and grouping traces that share the
same relative likelihoods all gave the same weights, differing by at
most \(6.6\times10^{-16}\), with terminal log-likelihoods within
\(7.3\times10^{-12}\). Second, the EM answer must match a completely
separate optimizer. A projected-gradient solver that never calls
ProteoEM's E-step agreed with the EM weights to \(3.78\times10^{-9}\)
and with an independent evaluation of the objective to
\(7.28\times10^{-12}\), in two full-rank cases (\(K=3,N=800\) and
\(K=8,N=6{,}000\)). Every multi-seed fit converged. These checks cover
clean, identifiable cases with a single best answer; they say nothing
about fits that are unidentifiable, on a boundary, or misspecified.

\textbf{Faster EM with equivalence classes.} ProteoEM speeds up the EM
by grouping molecules that carry the same evidence into one class with a
count, then iterating over classes rather than over molecules. This is
the same equivalence-class device RNA-seq uses to group reads that share
an alignment profile. Two molecules carry the same evidence when the EM
would split their counts among the candidates in the same proportions,
and that sameness can be recognized at three levels, each looser than
the one before. All three return the identical fitted composition,
because each added level only discards detail the assignment never uses;
they differ only in how much work they save.

The first level groups molecules with an identical raw trace, meaning
the same call in every cycle and the same pattern of missing calls. For
one million simulated molecules this alone gives 482,668 classes,
because only 32 of the 768 states are present and their traces recur.
The second level uses the fact that each logical probe is applied three
times with the same binding rate, so a probe's contribution to the
likelihood depends only on how many of its three calls were positive and
how many negative, and is unchanged by their order. For one probe, the
call patterns \((+,+,-)\) and \((+,-,+)\) carry the same evidence, two
positives and a negative. Grouping molecules with the same per-probe
counts gives 218,829 classes.

The third level drops the calls that cannot tell one candidate from
another. Two of the twelve are pan-tau probes that bind every candidate
at the same rate, so their call is equally likely whichever state a
molecule came from: it confirms the molecule is tau, but says nothing
about which tau it is. A call like this moves every candidate's
likelihood together, so it makes no difference once the EM weighs the
candidates against each other; molecules that agree on the other ten
probes then merge even when these two pan-tau calls differ. This gives
138,287 classes, a sevenfold reduction over the raw molecules.

Grouping is the only compression available here, because the affinity
likelihood matrix is dense: each probe binds with nonzero error rates,
so every trace keeps some likelihood under every candidate. Each class
still spans all 768 candidates, so even the most compressed E-step holds
106.2 million entries, against the 5.99 million stored links in the
29.45-million-fragment RNA-seq example, whose matrix is sparse (Table
1). The full breakdown is in Supplementary Results S4.1.

\clearpage

\textbf{Table 1. How grouping shrinks the E-step.} Every ProteoEM row
uses the same seeded synthetic design (\(C=36\) cycles, \(J=12\) logical
probes applied three times, \(K=768\) reference states, 32 present, 2\%
independent missing calls); the three one-million rows are the same
molecules grouped by three progressively looser rules. \(N\) is the
number of molecules, \(T\) the number of distinct groups the EM actually
scores, \(N-T\) the molecules that repeat an existing group, and \(N/T\)
the fold reduction. The last column is the work per EM iteration:
affinity data scores every group against all \(K\) candidates, so it is
dense (\(T\times K\) entries), while the RNA-seq row (the
29.45-million-fragment example in Table 2 of Zakeri et al. {[}21{]}) is
sparse and counts only stored class-transcript links, which is why 29
million RNA-seq reads need fewer entries than one million affinity
traces.

\begingroup
\scriptsize
\centering
\begin{tabular}{@{}>{\raggedright\arraybackslash}p{3.7cm}rrrrr@{}}
\toprule
Dataset and grouping rule & $N$ & $T$ & Copies $N-T$ & Fold $N/T$ & EM entries/iteration \\
\midrule
ProteoEM 5k-trace check: identical raw trace/mask & 5,000 & 4,339 & 661 & 1.15 & 3.33M dense \\
ProteoEM 1M: identical raw trace/mask & 1,000,000 & 482,668 & 517,332 & 2.07 & 370.7M dense \\
ProteoEM 1M: repeated-probe sufficient counts & 1,000,000 & 218,829 & 781,171 & 4.57 & 168.1M dense \\
ProteoEM 1M: relative trace-likelihood class & 1,000,000 & 138,287 & 861,713 & 7.23 & 106.2M dense \\
RNA-seq: compatibility class & 29,447,710 & 438,393 & 29,009,317 & 67.2 & 5.99M sparse \\
\bottomrule
\end{tabular}
\endgroup

\textbf{The value of keeping the full likelihoods.} ProteoEM weighs each
trace by its full calibrated likelihood. A simpler method would throw
that away, either by assigning each molecule to its single best origin
(top-likelihood counting) or by turning the likelihoods into yes/no
compatibility (binary-profile EM). Top-likelihood counting stands in for
hard decoding of the kind PrISM uses {[}13{]}; the fractional EM of IMaP
{[}15{]} is itself weighted, like ProteoEM. We compared all three on the
correctly specified model. Keeping the full likelihoods mattered. We
report accuracy as total-variation error (TV), half the total absolute
difference between the estimated and true composition, so 0 is an exact
match and 1 is complete disagreement. Weighted EM's error fell as the
sample grew, reaching 0.013 at \(N=20{,}000\), while the two reduced
methods stayed near 0.19 and 0.21--0.24 across sample sizes (Table 2).
Figure 6 shows why, candidate by candidate: weighted EM tracks the
present proteoforms closely, whereas the reductions both misestimate
them and spread mass onto proteoforms that are absent. The binary method
is not wrong in principle. When probes are noise-free it is exact, and
ProteoEM's EM reproduced the analytical answer
\((21/50,9/50,4/15,2/15)\), matching an independent incidence-matrix EM
to \(2.3\times10^{-13}\). Once the probes are noisy, thresholding throws
information away: in a six-origin, eight-probe sweep, weighted EM
reached \(0.0149\pm0.0040\) error against the best threshold's
\(0.0243\pm0.0049\) (Supplementary Results S4.2, Table S6).

\textbf{Table 2. Synthetic abundance error.} Entries are mean \(\pm\)
sample standard deviation of total-variation error across three seeds.
All runs used 768 reference origins, 32 nonzero generating origins, 36
cycles, oracle probe rates, and 2\% independent missing calls. These
favorable, correctly specified simulations test the implementation and
the consequence of discarding the likelihood values; they do not
establish expected assay performance or general superiority under
calibration error, dependence, censoring, or candidate-set
incompleteness.

{\def\LTcaptype{none} % do not increment counter
\begin{longtable}[]{@{}
  >{\raggedleft\arraybackslash}p{(\linewidth - 6\tabcolsep) * \real{0.2500}}
  >{\raggedleft\arraybackslash}p{(\linewidth - 6\tabcolsep) * \real{0.2500}}
  >{\raggedleft\arraybackslash}p{(\linewidth - 6\tabcolsep) * \real{0.2500}}
  >{\raggedleft\arraybackslash}p{(\linewidth - 6\tabcolsep) * \real{0.2500}}@{}}
\toprule\noalign{}
\begin{minipage}[b]{\linewidth}\raggedleft
Accepted molecules
\end{minipage} & \begin{minipage}[b]{\linewidth}\raggedleft
Weighted EM
\end{minipage} & \begin{minipage}[b]{\linewidth}\raggedleft
Top-likelihood counting
\end{minipage} & \begin{minipage}[b]{\linewidth}\raggedleft
Binary-profile EM
\end{minipage} \\
\midrule\noalign{}
\endhead
\bottomrule\noalign{}
\endlastfoot
1,000 & 0.0699 \(\pm\) 0.0103 & 0.1959 \(\pm\) 0.0074 & 0.2404 \(\pm\)
0.0075 \\
5,000 & 0.0310 \(\pm\) 0.0058 & 0.1882 \(\pm\) 0.0155 & 0.2127 \(\pm\)
0.0108 \\
20,000 & 0.0129 \(\pm\) 0.0014 & 0.1909 \(\pm\) 0.0090 & 0.2134 \(\pm\)
0.0082 \\
\end{longtable}
}

\begin{figure}
\centering
\includegraphics[width=1\linewidth,height=\textheight,keepaspectratio,alt={Weighted EM recovers proteoform abundances; reduced comparators do not. One correctly specified fit at N=20\{,\}000 accepted traces (seed 7) compares weighted EM against two baselines that force a hard yes/no: top-likelihood counting and binary-profile EM. (A) Estimated versus true abundance for the 32 present proteoforms (log-log; the dashed line is exact agreement). Weighted EM tracks the truth to a median 4\% error, the reductions to 21\% and 25\%. (B) Total abundance the same fits place on the 736 proteoforms that are truly absent; weighted EM leaves 0.003 there, while the reductions spread 0.18 and 0.22 of the estimated composition across proteoforms that are not present. The three total-variation errors for this single seed are consistent with the three-seed N=20\{,\}000 row of Table 2. All values are simulated.}]{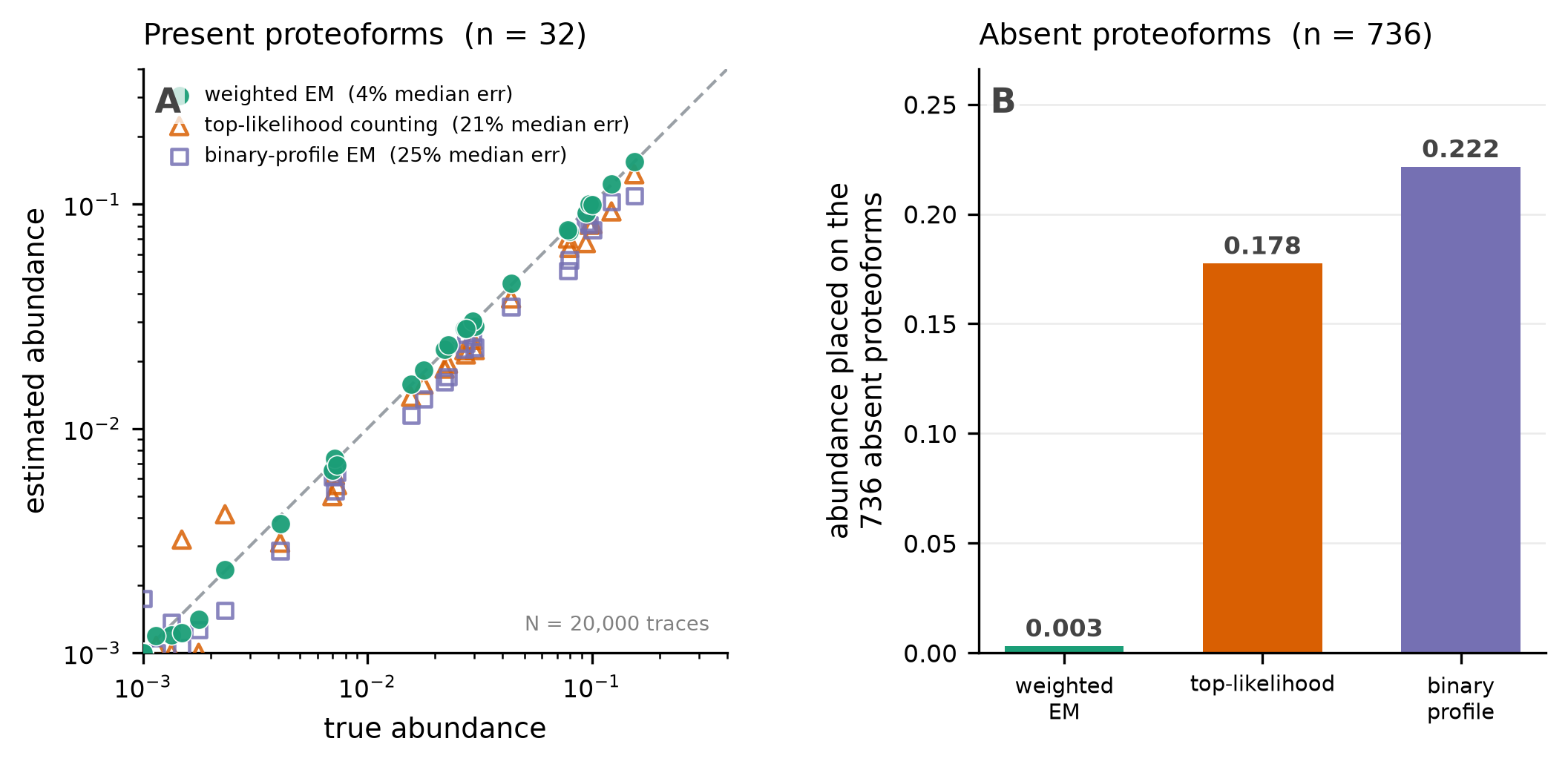}
\caption{\textbf{Weighted EM recovers proteoform abundances; reduced
comparators do not.} One correctly specified fit at \(N=20{,}000\)
accepted traces (seed 7) compares weighted EM against two baselines that
force a hard yes/no: top-likelihood counting and binary-profile EM. (A)
Estimated versus true abundance for the 32 present proteoforms (log-log;
the dashed line is exact agreement). Weighted EM tracks the truth to a
median 4\% error, the reductions to 21\% and 25\%. (B) Total abundance
the same fits place on the 736 proteoforms that are truly absent;
weighted EM leaves 0.003 there, while the reductions spread 0.18 and
0.22 of the estimated composition across proteoforms that are not
present. The three total-variation errors for this single seed are
consistent with the three-seed \(N=20{,}000\) row of Table 2. All values
are simulated.}
\end{figure}

The weighted EM phase ran in \(0.190\pm0.072\) s at \(N=1{,}000\),
\(0.939\pm0.238\) s at \(5{,}000\), and \(2.867\pm0.526\) s at
\(20{,}000\) on a 10-core Apple M5 (three seeds per size); these
characterize this reference implementation and say nothing about
production throughput. Runtime, peak memory, and the scaling audit
through one million simulated molecules are reported in Supplementary
Results S4.3.

\subsection{3.3 Behavior under model
violations}\label{behavior-under-model-violations}

A real experiment will not obey the model exactly, and we wanted to know
which of its assumptions we can break and still trust the abundances.
Four kinds of break matter most here. Calls go missing, and not always
at random. An instrument may drop positive calls more often than
negative ones, for example when a bright spot saturates. The emission
matrix \(Q\) is measured in separate control experiments and is never
exact. Some proteoforms share all their features, so no probe can tell
them apart. And a real sample will contain molecules whose true
proteoform is missing from the reference list. For each break we asked
the same question: can we ignore it, and if not, how far does it push
the answer?

\begin{figure}
\centering
\includegraphics[width=1\linewidth,height=\textheight,keepaspectratio,alt={Which model violations bias the abundance estimate. Mean total-variation abundance error (± sample SD across twenty seeds, 10,000 accepted traces each) under four departures from the fitting assumptions, against the ignorable-missingness baseline (dashed line). Group-level total variation is used throughout so the emission-equivalent case is scored at the resolution it supports. Miscalibrated Q and emission-equivalent proteoforms leave the error at baseline (green); informative missingness and an origin absent from the reference set bias the estimate (orange), by roughly 1.6-fold and 11-fold. Point estimates from twenty seeds; bootstrap intervals not computed. All values are simulated.}]{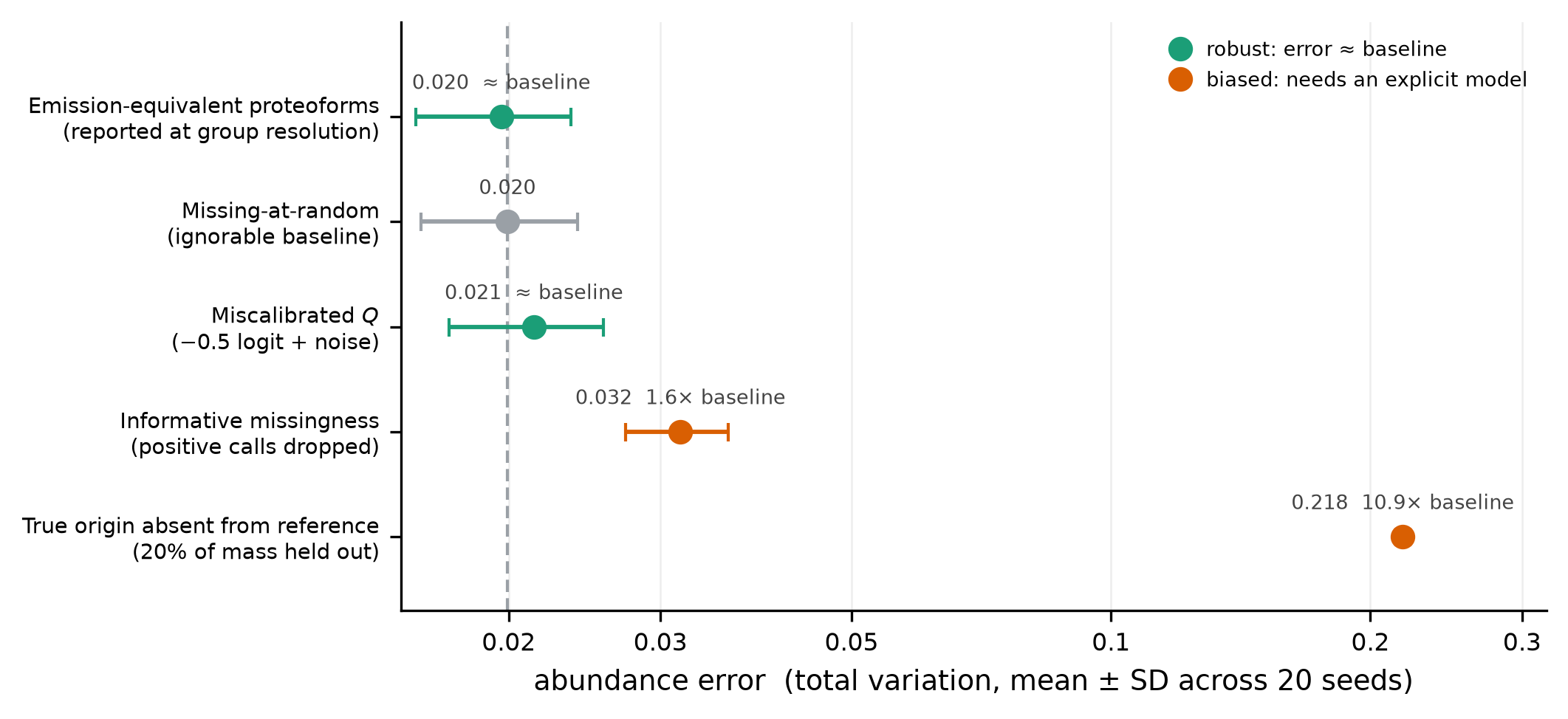}
\caption{\textbf{Which model violations bias the abundance estimate.}
Mean total-variation abundance error (± sample SD across twenty seeds,
10,000 accepted traces each) under four departures from the fitting
assumptions, against the ignorable-missingness baseline (dashed line).
Group-level total variation is used throughout so the
emission-equivalent case is scored at the resolution it supports.
Miscalibrated \(Q\) and emission-equivalent proteoforms leave the error
at baseline (green); informative missingness and an origin absent from
the reference set bias the estimate (orange), by roughly 1.6-fold and
11-fold. Point estimates from twenty seeds; bootstrap intervals not
computed. All values are simulated.}
\end{figure}

\textbf{The cost of each model violation.} We broke one assumption at a
time in the simulation, refit the abundances, and measured how far the
answer drifted from the known truth. Figure 7 lines up the four tests
against a baseline where nothing is broken. All fits converged. Two of
the breaks barely moved the answer; two moved it a lot.

The two harmless breaks were a wrong calibration and a pair of
proteoforms no probe can tell apart. When we fed the EM the wrong probe
rates (each shifted by \(-0.5\) on the logit scale, plus noise), the
answer hardly changed, because a shift that hits every candidate the
same way cancels out when the EM weighs candidates against each other.
This covers only one kind of miscalibration; a larger or lopsided error,
or probes whose calls are correlated, could still cause trouble, and we
have not tested those. In the second case we switched off the one probe
that separates the 3R and 4R proteoforms. The EM did not guess between
them. It reported their combined total instead, and that total was as
accurate as the baseline. The EM correctly reports two proteoforms as a
single group when no probe can separate them.

The two damaging breaks were informative missing data and a missing
candidate. If positive calls go missing more often than negative ones
(here 20\% against 2\%) and the EM still treats the gaps as random, the
estimate is biased, because the pattern of missing calls carries
information that the EM throws away; handling it needs a model of why
calls go missing. The worst case was a molecule whose true proteoform is
not in the reference list at all. We held out eight proteoforms that
together made up a fifth of the sample and asked the EM to explain those
molecules using only the ones that remained. It spread their weight
across the wrong candidates, and the extra weight did not land on any
single proteoform (each was off by only \(0.0006\) on average), so no
single wrong candidate stands out. The affected molecules do stand out
individually. With its true proteoform absent, a held-out molecule finds
no good match, so the per-molecule confidence ProteoEM already reports
runs much lower than for a molecule that belongs, a mean top posterior
of 0.63 against 0.99. Ranking molecules by that confidence separated
held-out from in-list molecules with an AUROC of 0.973 across twenty
seeds. The missing candidate still biases the abundances, but the
affected molecules can be flagged one at a time and set aside for a
wider search. The flag needs its own threshold. A cutoff at 0.80 still
passes about a third of the held-out molecules.

These are twenty-seed point estimates; the whiskers in Figure 7 show the
seed-to-seed spread. Bootstrap confidence intervals and several more
stress tests are planned (Supplementary Results S4.5).

\subsection{3.4 Observation yield and source
composition}\label{observation-yield-and-source-composition}

ProteoEM estimates the composition of the molecules it analyzes, but
those molecules are a biased subset of the source. Before a molecule can
be counted, it has to clear two hurdles. First it must make it onto the
chip at all, surviving sample handling and getting captured at a spot
where probes can read it. Then its trace has to pass the retention rule,
the filter the assay uses to discard spots that carry no usable signal;
a common rule keeps a trace only if at least one probe called positive.
The chance a molecule makes it onto the chip is its recovery, and the
chance its trace then passes the retention rule is its visibility. The
chance it is counted at all is these two multiplied, which we call the
effective observation yield:

\[
e = r \times v = \text{recovery} \times \text{visibility}.
\]

Here \(e\) is the yield, \(r\) the recovery, and \(v\) the visibility. A
proteoform's yield is the fraction of its source molecules that clear
both hurdles and get counted. Both factors can differ from one
proteoform to the next, and that difference is what biases the count.

Visibility shows this most clearly. Take two proteoforms present in
equal amounts. One is bound by several probes, so it almost always
lights up somewhere and passes an at-least-one-positive gate nearly
every time. The other is bound by only one probe, so it often comes back
all-negative, looks like an empty spot, and is thrown away. The first is
then counted far more often than the second, and the surviving tally no
longer matches the source.

Recovering the source composition then takes two steps. The first gets
the makeup of the kept molecules right. That is more than tallying them,
because the gate changes what the surviving traces look like. It
discards every all-negative trace, so a proteoform that usually reads
all-negative survives only through its uncommon traces that happen to
fire a probe, and in the kept set it looks more active than it typically
is. The analysis has to account for that as it weighs each ambiguous
trace. The second step scales that kept makeup back to the source,
dividing each proteoform by how often it survives. If one survives 9
times in 10 and another only 5 in 10, each kept molecule of the rarer
one is counted for nearly twice as much as each kept molecule of the
common one. Applied to the two proteoforms above, that rescaling returns
them to the equal amounts the sample held. We tested both steps on a
two-origin, 50:50 mixture under three retention rules: retain every
trace; retain any trace with at least one positive call; and a synthetic
gate requiring both an N-terminal and a C-terminal anchor.

\textbf{Getting the accepted mixture right.} When every trace is kept,
nothing is skewed. Both origins are recorded whether they carry 12 probe
opportunities or 3, and the 50:50 mixture came back with a
total-variation error of about 0.003. The any-positive gate is where the
skew shows up. One origin passed the gate 93\% of the time and the other
only 49\%, so the 50:50 source looked like 66:34 among the kept traces.
To undo this, ProteoEM divides each candidate's likelihood by how often
that candidate passes the gate, its visibility \(v_k\). Because \(v_k\)
is computed from the calibrated \(Q\) and the known retention rule
rather than fit to the mixture, this gate correction applies to real
data whenever \(Q\) is calibrated and is not limited to simulations.
That recovered the accepted mixture to 0.0025 error; ignoring the gate
and fitting the raw likelihoods gave 0.049, about twenty times worse.

\textbf{Getting back to the source.} The accepted mixture is still not
the source. To recover the original sample we apply the effective yield
\(e=rv\) once. Visibility comes from \(Q\); recovery \(r\) does not, and
in a real experiment it would come from separate calibration, for
example spike-in standards of known composition. So the gate half of the
correction is available now, while the recovery half is what a real
study still has to measure. With the true yield the 50:50 source came
back to about 0.003 error under all three keep-rules. Shortcuts for the
yield do not work. Assuming equal yield, or scaling by probe count, by
protein length, or by a guessed yield, gave errors of 0.16, 0.18, 0.29,
and 0.04. These are worked counterexamples; they do not rank practical
proxies. The probe-count ratio matched the true yield in the N+C case
only by coincidence. Filtering by decode confidence is not a substitute
either. Keeping only molecules decoded at 90\% confidence retained 55\%
of them and shifted the two origins to 63:37. That filtered set can look
accurate against its own filtered truth, but it is not an unbiased
estimate of the source, because the confidence filter itself chose which
molecules to count.

These are correctly specified two-origin simulations. They check the
bookkeeping and the implementation. They do not measure real recovery,
do not establish how the assay would perform, and, because they use an
oracle \(Q\), do not test how calibration error would propagate into
both the likelihoods and the computed visibility \(v_k\). Figure S2
shows the accepted- and source-composition errors in full, and Table 3
summarizes the yield-benchmark errors.

\textbf{Table 3. Observation-yield benchmark: total-variation error on a
two-origin, 50:50 source.}

{\def\LTcaptype{none} % do not increment counter
\begin{longtable}[]{@{}
  >{\raggedright\arraybackslash}p{(\linewidth - 2\tabcolsep) * \real{0.4286}}
  >{\raggedleft\arraybackslash}p{(\linewidth - 2\tabcolsep) * \real{0.5714}}@{}}
\toprule\noalign{}
\begin{minipage}[b]{\linewidth}\raggedright
Estimate
\end{minipage} & \begin{minipage}[b]{\linewidth}\raggedleft
TV error
\end{minipage} \\
\midrule\noalign{}
\endhead
\bottomrule\noalign{}
\endlastfoot
Accepted composition, gate ignored & 0.049 \\
Accepted composition, gate-conditioned & 0.0025 \\
Source composition, with oracle yield \(e\) & 0.0028 \\
Source composition, via yield shortcuts (equal / probe-count / length /
guess) & 0.16 / 0.18 / 0.29 / 0.04 \\
\end{longtable}
}

Five-seed selection and inverse-yield checks (gate-conditioned
\(0.00251\pm0.00204\); oracle-yield source \(0.00278\pm0.00227\)); the
all-trace design with no gate recovers the source to about 0.003.
Shortcut and gate-ignored entries are single-configuration point
estimates.

\section{4. Discussion}\label{discussion}

ProteoEM is an open-source computational framework for estimating
proteoform abundances from iterative affinity-probe measurements. It
adapts the expectation-maximization approach used for transcript
quantification in RNA sequencing to affinity-based proteomics, where
each single-molecule trace is compatible with several candidate
proteoforms. ProteoEM holds a pre-calibrated emission matrix fixed and
separate from abundance estimation, keeping calibration and
quantification independent and auditable. It keeps each trace's full
likelihood over all candidates, which preserves the evidence carried by
imperfect positive and negative measurements. It also separates the
composition of the molecules it observes from the composition of the
source sample, correcting for how readily each proteoform is detected.
In the simulations, retaining the full likelihoods recovered the
generating composition to a total-variation error of 0.013 at 20,000
traces, where methods that discard the full likelihoods stayed at 0.19
and above, and reporting proteoforms only at the resolution the panel
supports kept indistinguishable species from being split arbitrarily.

The simulations in this study evaluate the estimator under controlled
conditions with known ground truth. Drawing samples from the generative
model lets us measure how close an estimate comes to the truth and turn
each modeling assumption on or off in isolation, from missing calls and
calibration error to proteoforms that share all measured features and
true proteoforms absent from the reference set. Shared miscalibration
and unresolvable proteoforms were tolerated at the group level.
Informative missing data and omitted proteoforms biased the estimate.
The method comparison also favors weighted EM. The heavy-tailed
composition leaves many proteoforms at low abundance, where hard
decoding fails most, and a flatter composition would narrow the gap.
Because the idealized emission matrix has no real calibration error,
recovery variation, or correlated noise, these results describe the
estimator and its bookkeeping. They do not show how an experimental
platform would perform.

ProteoEM builds on two published methods for iterative single-molecule
affinity mapping. PrISM set out a proteome-scale identification
framework that assigns each molecule to its single best-scoring protein
{[}13{]}, while the tau IMaP study demonstrated the experimental assay
and used EM to count proteoform groups fractionally, with binding rates
estimated jointly from the sample {[}15{]}. Like IMaP, ProteoEM
apportions each ambiguous molecule by EM rather than forcing it to one
identity as PrISM does. It differs from IMaP in holding the emission
matrix fixed from external calibration rather than learning it from the
mixture, and in making the correction from observed to source
composition and the group-resolution rule explicit. Fixing the emission
matrix also makes the objective concave for a specified likelihood,
which allows the fit to be checked against an independent optimizer.
Analyses on commercial platforms are often proprietary. An open
implementation lets others check its assumptions, reproduce its results,
and build on it.

As in any latent-mixture model, the resolution ProteoEM can achieve is
set by the information in the measurements. Proteoforms with distinct
affinity profiles separate readily, while those with identical profiles
under the panel cannot be told apart by any amount of data and are
identifiable only through their combined abundance. ProteoEM handles
this explicitly, detecting proteoforms with identical emissions and
reporting them as one observable group rather than splitting their
abundance arbitrarily. In the simulations, a single discriminating probe
was enough to separate the 2N4R and 2N3R tau isoforms, and without it
the two collapsed into one correctly quantified group. Probe diversity
and panel design therefore set the ceiling on the proteoform resolution
that computation can deliver. The same formulation extends to a
proteome-wide reference, where the limiting factor becomes candidate-set
completeness, since a true origin missing from the reference has its
abundance redistributed across the candidates that remain.

The next step is to apply ProteoEM to experimental data. This means
measuring \(Q\) on standards of known identity, and measuring recovery
with spike-in standards. If the calibration is incomplete, \(Q\) could
be adjusted to fit the sample while staying close to its calibrated
values. This would sit between ProteoEM, which fixes \(Q\), and IMaP,
which learns \(Q\) from the sample. It would let \(Q\) follow how the
probes behave in the real sample. The cost is that the fit would no
longer have a single best answer, and errors in \(Q\) could no longer be
told apart from errors in abundance. A proteome-wide reference raises
two further problems. First, it can never be complete. An extra
``unknown'' component could collect the molecules that match no
candidate well, so their abundance is not spread onto real proteoforms.
The per-molecule confidence already flags these molecules. Second, a
much larger candidate set makes the computation heavier and leaves more
proteoforms that no probe can tell apart. It would need sparse
representations or validated approximations, and its results would rely
more on group-level reporting and on checks for proteoforms that are
nearly indistinguishable.

The abundance estimates are point estimates at present. Intervals would
show which estimates are firm and which are uncertain {[}22{]}, and
whether a difference between two proteoforms, or between two samples, is
larger than sampling noise. A nonparametric bootstrap over accepted
traces would capture sampling variability {[}23{]}. An outer bootstrap
over the calibration data would propagate uncertainty in \(Q\). A
Bayesian treatment with a Dirichlet prior on the composition would give
credible intervals directly {[}24{]}. Holding \(Q\) fixed makes the
likelihood concave, which keeps this tractable. A hierarchical prior on
\(Q\) could extend it to calibration uncertainty. Because many
proteoforms are estimated at exactly zero, the composition lies on the
boundary of the simplex, where resampling and Bayesian intervals are
more reliable than asymptotic approximations. These are directions we
leave for future work.

ProteoEM turns imperfect affinity traces into proteoform abundance
estimates while keeping calibration, resolution limits, and per-molecule
uncertainty explicit. We release it openly as a foundation for weighted
estimation of proteoform abundance from single-molecule affinity
measurements.

\section{Data and code availability}\label{data-and-code-availability}

ProteoEM is available as open-source software (version 0.1.1) under the
MIT license at \url{https://github.com/narayananr/proteoem}, with an
archived release deposited at Zenodo
(\href{https://doi.org/10.5281/zenodo.22721009}{doi:10.5281/zenodo.22721009}).
The repository includes the Python package, 100 tests, a command-line
smoke benchmark, a dependency lock file, and the scripts that regenerate
every benchmark and figure in this manuscript. Each benchmark writes a
self-contained archive under \texttt{outputs/} (empirical,
observation-yield, independent-correctness, and trace-class-scale), with
per-run and summary tables, environment metadata, and source hashes;
Figure S2 is generated from the observation-yield tables. Beyond
computing likelihoods from \(Q\), ProteoEM also accepts a precomputed
per-trace likelihood matrix as input; the validity condition for a
supplied likelihood is given in Supplementary Methods S2.3.
Supplementary Material (Supplementary Note S1, Supplementary Methods
S2--S3, Supplementary Results S4, and Supplementary Software)
accompanies this manuscript. All data reported here are simulated and
regenerable from the archived configurations; none derive from Nautilus
assays, data, or calibration.

\section{Author contributions}\label{author-contributions}

Narayanan Raghupathy conceived the study, implemented the software,
designed and performed the simulations and benchmarks, and wrote the
manuscript, using generative-AI tools under the author's direction as
detailed in Use of AI.

\section{Competing interests}\label{competing-interests}

N.R. is the founder of Tensoromics LLC, a computational biology
consulting company that applies and develops AI and machine-learning
methods for computational biology in support of drug discovery. The
author declares no other competing interests.

\section{Funding}\label{funding}

This research received no external funding.

\section{Acknowledgements}\label{acknowledgements}

The author thanks the developers of the open-source scientific Python
ecosystem on which ProteoEM is built.

\section{Use of AI}\label{use-of-ai}

\textbf{AI-assisted methodology, implementation, analysis, and writing.}
The author developed the scientific problem formulation and proposed the
methodological solution, and directed all use of generative AI tools
throughout the project.

Under the author's direction, Claude Opus 4.8 (Anthropic) and Sol
(OpenAI) helped enhance the expectation-maximization model for
affinity-trace data, extend the author's existing code to implement it,
build the software package and its tests, write the figure-rendering
scripts, and run the synthetic benchmarks. Those two models, together
with Gemini 3.1 Pro (Google), helped draft, edit, and restructure the
manuscript and supplement, including reference formatting.

The author reviewed, modified, tested, and validated all AI-assisted
contributions (code, figures, analyses, and text) and is solely
responsible for the final content, including the underlying methodology,
computational implementation, analyses, and scientific conclusions.

\section{References}\label{references}

\begingroup
\footnotesize

\protect\phantomsection\label{refs}

\endgroup

\end{document}